\documentclass[12pt]{article}
\usepackage{graphicx,amsmath}

\newlength{\figsize}
\begin{document}

\begin{titlepage}

\begin{tabbing}
\` Oxford preprint OUTP-03-19P \\
\end{tabbing}
 
\vspace*{0.7in}

\begin{center}
{\large\bf The high temperature phase transition in SU(N) gauge theories \\ }
\vspace*{0.5in}
{Biagio Lucini, Michael Teper and Urs Wenger\\
\vspace*{.2in}
Theoretical Physics, University of Oxford,\\
1 Keble Road, Oxford, OX1 3NP, U.K.\\
}
\end{center}

\vspace*{0.75in}

\centerline{\it Dedicated to the memory of Ian Kogan, friend and colleague.}

\vspace*{0.75in}

\begin{center}
{\bf Abstract}
\end{center}

We calculate the continuum value of the deconfining temperature
in units of the string tension for SU(4), SU(6) and SU(8) gauge 
theories, and we recalculate its value for SU(2) and SU(3). We 
find that the $N$-dependence for $2\leq N\leq 8$ is well fitted by
$T_c/\surd\sigma = 0.596(4) + 0.453(30)/N^2$, showing a rapid 
convergence to the large-$N$ limit. We confirm our 
earlier result that the phase transition is first order 
for $N\geq 3$ and that it becomes stronger with increasing $N$.
We also confirm that as $N$ increases the finite volume 
corrections become rapidly smaller and the phase transition
becomes visible on ever smaller volumes. We interpret the latter 
as being due to the fact that the tension of the domain wall 
that separates the confining and deconfining phases increases 
rapidly with $N$. We speculate on the connection to 
Eguchi-Kawai reduction and to the idea of a Master Field.

\end{titlepage}

\setcounter{page}{1}
\newpage
\pagestyle{plain}

\section{Introduction}
\label		{intro}

In a recent letter
\cite{letterTc}
we presented lattice calculations of the finite temperature 
deconfining transition in SU(4) and SU(6) gauge theories. 
These calculations showed that the phase transition is first 
order and becomes more strongly so as $N$ grows. Our continuum
extrapolations, taken together with older SU(2) and SU(3)
results, showed that the $O(1/N^2)$ corrections
to the $N=\infty$ limit appear to be small, even for SU(2).
Finally, our finite volume studies showed that the
finite volume corrections decrease as $N$ grows and we 
speculated on the simple picture that this suggested
for SU($N=\infty$).

In this paper we will provide more accurate and more extensive
calculations that confirm and make more precise our earlier 
study. For SU(4) and SU(6) we extend the range of
lattice spacings at which we perform the calculations, thus
allowing more reliable and precise continuum extrapolations.
In addition we perform some calculations in SU(8) to
strengthen our control of the $N\to\infty$ limit. We also
perform some SU(2) and SU(3) calculations, both to
check that our methods produce results in agreement with
older (but still accurate) calculations and to investigate
the crucial volume dependence in more detail. In addition
we perform dedicated string tension calculations for all the
above SU($N$) groups, generally at a value of the lattice
coupling very close to the phase transition. This again
improves the accuracy of our calculation and, as a by-product,
leads to significant changes to the older continuum extrapolations
of the SU(2) and SU(3) deconfining temperatures.

The plan of the paper is as follows. In the next section
we briefly summarise the details of the lattice setup.
We then describe how we will identify the location and
nature of the phase transition. We follow this with
the details of our lattice results. We focus on 
$T_c/\surd\sigma$, the value of the deconfining temperature 
in units of the string tension, on the order of the transition,
on the (Euclidean) latent heat, where first order, and on the 
$N$-dependence  of the finite volume corrections. We finish
with some conjectures about the nature of physics at $N=\infty$,
and with some conclusions.

We have, in addition, studied how topological fluctuations
and ($k$-)string masses vary across the phase transition,
how the bulk transition and the thermodynamic latent heat vary 
with $N$, and other aspects of the transition. These, together
with more details about the technical aspects of our
calculations, will be presented in a later publication
\cite{laterTc}.
For some comments on the interest of SU($N\to\infty$)
gauge theories
\cite{largeN},
some references to relevant lattice calculations,
and some speculations 
\cite{pisarski}
on the nature of the deconfining transition, we refer the 
reader to our earlier paper
\cite{letterTc}.
%

\section{Lattice setup}
\label{section_lattice}

Our lattice calculations are entirely conventional. We use 
hypercubic, periodic lattices with SU($N$) matrices, $U_l$,
assigned to the links $l$.  We use the usual plaquette action 
\begin{equation}
S = \beta \sum_{p}\{1-{1\over N}{\mathrm {ReTr}} U_p\},
\label{B1}
\end{equation}
where $U_p$ is  the ordered product of the SU($N$) matrices 
around the boundary of the plaquette $p$, and which appears 
in the Euclidean path integral as $e^{-S}$.
This lattice action becomes the usual Yang-Mills action with
\begin{equation}
\beta = {{2N} \over {g^2}}
\label{B2}
\end{equation}
in the continuum limit. The coupling $g^2$ is the bare coupling
and, on a lattice of spacing $a$, may be regarded as a running 
coupling, $g^2(a)$, 
defined at the length scale $a$ at which the action is defined. 
So to decrease  $a$ we decrease $g^2(a)$ and hence increase $\beta$ 
(once $a$ is small enough). As we vary $N$ we expect 
\cite{largeN},
that we will need to keep constant the 't Hooft coupling, $\lambda$,
and its inverse, $\gamma$,
\begin{equation}
\lambda(a) \equiv g^2(a) N  
\ \ ; \ \ \ \ \
\gamma \equiv {1\over{\lambda}} ={{\beta}\over{2N^2}}
\label{B2b}
\end{equation}
for a smooth large-$N$ limit. Non-perturbative calculations
\cite{blmt-kstring,blmt-glue}
support this expectation.

A periodic lattice whose size is $L^3 L_t$ with $L \gg L_t$ 
may be used to calculate thermal averages of the field theory
at a temperature
\begin{equation}
a(\beta) T = {1\over{L_t}}.
\label{B3}
\end{equation}
We can therefore study the deconfining transition by fixing 
$L_t$ and varying $\beta$ so that $T$ passes through $T_c$. 
For simplicity we shall refer to the 3+1 dimensional system
as being at the temperature $T$ given by eqn(\ref{B3}),
although  this can be deceptive (see below) and one needs to remember
that the D=3+1 system is no more than a theoretical framework for 
calculating thermal averages in a given three dimensional spatial 
slice at this value of $T$.

The simulations are performed with a combination of heat bath 
and over-relaxation updates, as described in 
\cite{blmt-kstring,blmt-glue}.
We update all the $N(N-1)/2$ SU(2) subgroups of the SU($N$)
link matrices.

\section{The phase transition}
\label{section_transition}

We begin with some general comments about the deconfining
phase transition and then we discuss how this phase
transition appears in a finite volume and how one can 
extrapolate to the desired $V=\infty$ thermodynamic limit. 
The discussion here will be heuristic, but can be formalised
through standard finite size scaling theory, both for first
\cite{FSS1}
and for second
\cite{FSS2}
order transitions. (We will summarise relevant aspects of
this approach in our forthcoming paper
\cite{laterTc}.)
%

\subsection{General remarks}
\label{subsection_general}

The deconfining phase transition turns out to be second order 
for SU(2) and first order for SU($N\geq 3$). For a second 
order transition we have a diverging correlation length
as $T\to T_c$ while for a first order transition
various quantities will suffer a discontinuity
at $T=T_c$ (including the correlation length which
in general will be finite). This clear-cut distinction
only holds in the $V=\infty$ thermodynamic limit. On
a finite volume, as in our numerical calculations, the
two types of transition can be confounded and so a careful 
finite volume study is obligatory. We shall describe 
later on in this Section how we analyse the two
kinds of phase transition.

We are of course interested in the continuum limit of
this phase transition. The continuum limit of the lattice
theory is described by a 2'nd order phase transition at 
$\beta=\infty$. This means that when we study the first order
deconfining transition on the lattice, it will always appear to 
be weak; for example the latent heat is $O(a^4)$
and therefore small. Even on the coarser lattice spacings 
that are close to the bulk transition, which is first order for 
larger $N$, its latent heat is typically 1/50'th of the latter's. This
is because the deconfining phase transition
is a physical transition that only affects fluctuations  on physical 
length scales, and leaves untouched the short distance
fluctuations that usually dominate the values of composite operators.
A typical lattice transition, like the bulk transition, affects 
fluctuations on all length scales.
So to judge whether the first order phase transition is strong or 
weak we should compare the latent heat to a physical quantity,
such as the $T=0$ gluon condensate.

The phase transition is characterised by the fact 
that for $T<T_c$ the confining string tension is non-zero, 
$\sigma\not= 0$, while for $T>T_c$ we have  $\sigma= 0$. To
calculate the string tension we calculate the free energy
of two static sources in the fundamental representation 
placed a distance $r$ apart, $F_{s{\bar s}}(r)$. 
Inserting such minimally coupled 
static sources into the action is equivalent to calculating
the correlator of two Polyakov loops 
$\langle l_p^{\dagger}(r) l_p(0) \rangle$ where
\begin{equation}
l_p(\vec{n}_s) = \prod_{n_t=1}^{n_t=L_t} U_0(\vec{n}_s,n_t).
\label{C1}
\end{equation}
(Here we introduce the notation, $U_\mu(n)$, for the
matrix on the link emanating from site $n$ in direction $\mu$.)
In the confining phase at temperature $T=1/aL_t$
\begin{equation}
e^{-{{F_{s{\bar s}}(r)}\over{T}}}
=
\langle l_p^{\dagger}(r=an_z) l_p(0) \rangle
\stackrel{r\to\infty}{\propto}
e^{- a^2\sigma(T) n_z L_t}
 \  \   \   \   \  \  \  \ T<T_c 
\label{C2}
\end{equation}
where, to be definite, we take correlations in the $z$-direction.
Since the operators $l_p$ are localised in a given $z$-slice, this
correlator can be expressed in terms of the eigenstates and eigenvalues
of the transfer matrix $T_z$ that translates us in the $z$-direction: 
\begin{equation}
\langle l_p^{\dagger}(r=an_z) l_p(0) \rangle
=
\sum_n |\langle vac |l_p^{\dagger} |n \rangle|^2 e^{-a m_n n_z}
\label{C3}
\end{equation}
where $T_z|n\rangle = e^{-am_n}|n\rangle$.
If the lightest state in the sum is $m_l$ then
\begin{equation}
\langle l_p^{\dagger}(r=an_z) l_p(0) \rangle
\stackrel{n_z\to\infty}{\propto}
e^{-a m_l n_z}
\label{C4}
\end{equation}
We see from eqns(\ref{C2},\ref{C4}) that 
\begin{equation}
a m_l =  a^2\sigma(T) L_t
\label{C4b}
\end{equation}
and the fact that it is $\propto L_t$ indicates 
that this state of the $z$-transfer matrix is the lightest flux loop 
that winds once around the $t$-torus. Once we have determined
that the interesting quantity is this lightest flux loop, we
can alter the operator in the correlator to make the calculation
more efficient. For example, we smear/block 
\cite{blmt-kstring,blmt-glue}
the links to have a better overlap onto the extended lightest state
and we take $p_x=p_y=0$ sums of these smeared Polyakov loops
so as to exclude explicitly the more energetic states with 
$\vec{p}\not= 0$. 

In the confining phase the Polyakov loop cannot have an overlap 
onto a non-winding state since otherwise we would lose
the factor of  $L_t$ in the exponent of eqn(\ref{C2}).
This constraint is enforced by a $Z_N \subset SU(N)$ symmetry
of the action and integration measure:
\begin{equation}
U_0(n_s,n_t) \to z U_0(n_s,n_t) 
\ \ \ \ \forall n_s
\label{C4c}
\end{equation}
where $z \in Z_N$ is any one of the $N$'th roots of unity
and $n_t$ is some fixed value that one can freely
choose. This symmetry ensures that  $\langle l_p \rangle =0$
since $l_p \to z l_p$ under the symmetry transformation.

In the deconfined phase $\sigma(T) = 0$ and we therefore lose 
the factor of $n_z$ in the exponent of eqn(\ref{C2}) and so,
in general,
$\lim_{r\to\infty}\langle l_p^{\dagger}(r) l_p(0) \rangle \not= 0$,
i.e.  $\langle  l_p  \rangle \not= 0$. In terms of the $Z_N$
symmetry described above, this is only possible if that
symmetry is spontaneously broken. This tells us that for $T>T_c$
there are $N$ degenerate deconfined vacua characterised by
$\langle  l_p  \rangle = c e^{i2k\pi/N} \ ; k=0,...,N-1$ for 
$c$ some real positive number. At very high $T$, where by
asymptotic freedom $g^2(T) \ll 1$, one can calculate the
tensions of the domain walls separating these vacua in
perturbation theory
\cite{cka_wallN}.
It is important to note that what we actually know is that this
$Z_N$ symmetry breaking occurs in the D=3+1
lattice system as one passes from $l_t\equiv aL_t > 1/T_c$
to  $l_t < 1/T_c$, and indeed in the continuum limit
of that system. It is not however at all clear that it is
actually a feature of the field theory at high $T$. That is to
say, no-one has (we believe) succeeded in showing that
this symmetry breaking is encoded in any thermal averages
in a way that displays the existence of such $Z_N$ bubbles
and domain walls in the hot field theory
\cite{smilga_zN}.
To avoid tedious circumlocution, we shall usually speak of our 
D=3+1 system as being at a temperature $T=1/aL_t$ although it is
actually a framework within which to calculate thermal 
averages in a given spatial slice, and so one must 
always be careful when ascribing the detailed dynamics of 
this 3+1 dimensional system to the hot field theory. 

In summary, the natural order parameter for the deconfining
phase transition is the average Polyakov loop, 
$\langle l_p  \rangle$, and the mass of the flux loop that
winds around the $t$-torus is the natural mass to focus upon,
since it is simply proportional to $\sigma (T)$.

\subsection{Characterising first order transitions}
\label{subsection_firstorder}

If the deconfining transition is first order, the finite-$T$ 
effective potential will have both a confining vacuum and $N$ 
degenerate deconfined vacua. At  $T=T_c$ all of these vacua
are degenerate. On a finite volume there will be a finite
range of $T$ around $T_c$ in which the system has a significant
probability to be in both confining and deconfined vacua, 
occasionally tunnelling between them. As we move away from
$T=T_c$, or $\beta=\beta_c$, the degeneracy lifts and
the confining/deconfined vacuum energies will split by
some $\Delta E$. The relative probabilities of the
two vacua are given by $e^{-\beta\Delta E}$. Since
$\Delta E \propto V$ we see that the finite volume
transition region is typically $\delta\beta \sim O(1/V)$. 
To calculate the $V=\infty$ value of $\beta_c$, and
hence of $T_c$, we must first make a choice of $\beta_c$
on a finite volume, so that we can extrapolate it.
A  sensible criterion will use a value of $\beta_c(V)$
that is in the transition region, and therefore we
generically can expect $\beta_c(V)=\beta_c(\infty)+O(1/V)$.
We shall see examples of this below.

In an infinite volume the average plaquette will jump across
the transition and the location of this jump defines for us
the (pseudo-)critical coupling $\beta_c(V=\infty)$ and the critical
temperature $a(\beta_c)T_c=1/L_t$. In a finite volume
the change will be continuous but rapid (if $V$ is not too small). 
A useful quantity to characterise this behaviour is the specific heat
$C(\beta)$:
\begin{equation}
{1\over{\beta^2}} C(\beta) 
\equiv  
{\partial\over{\partial\beta}} 
\langle \bar{u}_p  \rangle
=
{N_p}
\langle \bar{u}_p^2 \rangle 
- 
{N_p}
{\langle \bar{u}_p  \rangle }^2
\label{C5}
\end{equation}
where $u_p\equiv{\mathrm {ReTr}} U_p/N$ and  $\bar{u}_p$ is the 
average value of $u_p$ for a given lattice field, and $N_p=6L^3L_t$
is the total number of plaquettes. The maximum of the specific
heat is the maximum of $\partial\langle\bar{u}_p\rangle/\partial\beta$ 
(neglecting the weakly varying $\beta^2$ factor) and so the $\beta$ 
at which it occurs provides at least one sensible definition of $\beta_c(V)$.

Let $\bar{u}_{p,c}$ and $\bar{u}_{p,d}$ be the average plaquette
value for a lattice field in the confined and deconfined phase 
respectively, at $\beta=\beta_c(V)$. This value will fluctuate around
its average value, $\langle \bar{u}_{p,c} \rangle$ or
$\langle \bar{u}_{p,d} \rangle$, by an amount that is
$O(1/\surd N_p)$. As $V\to\infty$ we can neglect these fluctuations
and we can also neglect the very rare fields where the system is
tunnelling from one phase to another. In that case 
it is easy to see that the maximum of $C(\beta)$ occurs
when the lattice fields are equally likely to be in the
confined and deconfined phases, and in that case 
\begin{equation}
\lim_{V\to\infty} {1\over{\beta_c^2 N_p}} C(\beta_c)
=
{1\over{4}}
(\langle \bar{u}_{p,c} \rangle - \langle\bar{u}_{p,d}\rangle)^2
=
{1\over{4}} L^2_h
\label{C6}
\end{equation}
where $L_h$ is the latent heat per site of the phase transition and
$\beta_c$ denotes the value of $\beta$ at which the maximum 
of $C(\beta)$ occurs.  
(Note that this latent heat is that of the transition of the
D=3+1 system; it is related to, but differs from, the
latent heat of the finite-$T$ transition of the field theory.)
The $V$-dependence in eqn(\ref{C6}) 
\begin{equation}
\lim_{V\to\infty} C(\beta_c) \propto V
\label{C6b}
\end{equation}
is a direct consequence of the discontinuity in $\bar{u}_{p}$ 
and so is characteristic of a first order transition.

At a finite but large value of $V$,  $\bar{u}_{p}$ will fluctuate
around $\langle \bar{u}_{p} \rangle$ by an amount 
$O(1/\surd N_p) = O(1/\surd V)$, i.e. $\langle \bar{u}_p^2 \rangle 
- {\langle \bar{u}_p  \rangle }^2 = O(1/V)$. (We can neglect
the exponentially suppressed tunnelling fields.) It is then
easy to show that   $C(\beta_c^C)$ will receive
a correction that is $ O(1/V)$, as long as the fluctuations
in the confined and deconfined phases differ. Thus we should
expect the leading large-volume correction to both
$C(\beta_c^C)$ and $\beta_c^C(V)$ to be $O(1/V)$.

As we pointed out in Section~\ref{subsection_general}, the
natural order parameter for the deconfining transition
is the average Polyakov loop, $\langle l_p \rangle$. 
On a finite volume however,
$\langle l_p \rangle = 0 \ \forall T$ because of the tunnelling
between the $N$ deconfined vacua in the deconfined phase.
(This is simply the statement that one only  has 
spontaneous symmetry breaking in an infinite volume.)
To finesse this problem we follow usual practice and
replace $\langle l_p \rangle$
by $\langle |{\bar l}_p| \rangle$ where ${\bar l}_p$ is the 
average of the Polyakov loop over a single lattice field.
Clearly $\langle |{\bar l}_p| \rangle \not= 0 \ \forall T>T_c$. 
The price one pays is that now
$\langle |{\bar l}_p| \rangle \not= 0$ also for $T<T_c$.
However it is $O(1/\surd V)$ in the confining phase
and for a large volume this is a small price to pay.
We can use this to define a quantity analogous to $C(\beta)$, 
where we replace the plaquette with the Polyakov loop.
This is the normalised Polyakov loop susceptibility $\chi_l$
\begin{equation}
{{\chi_l}\over{V}} 
= 
\langle |{\bar l}_p|^2 \rangle - {\langle |{\bar l}_p| \rangle}^2.
\label{C8}
\end{equation}
Just as with the specific heat we can define $\beta_c(V)$
to be the value of $\beta$ at which $\chi_l$ has its maximum.
At finite $V$  the two definitions of $\beta_c(V)$ will differ 
by $O(1/V)$ terms, but will coincide in the thermodynamic limit.
(We shall refer to them as $\beta_c^\chi$ and  $\beta_c^C$
when we wish to distinguish them.)
Again we have the characteristic first order behaviour  
$\chi_l \propto V$ and corrections down by one power of $V$:
\begin{equation}
\lim_{V\to\infty} {{\chi_l}\over{V}} 
= 
c_0 + {{c_1}\over{V}} +\ldots
\label{C8b}
\end{equation}
Just as for the specific heat, the maximum of $\chi_l$
occurs where the system is equally likely to be in
the confined and deconfined phases (for $V\to\infty$).
One might think that a more natural criterion would
be one where the deconfined phase was $N$ times more
probable than the confined one (simply because of the
relative multiplicity of the former) and that it would
therefore have smaller $O(1/V)$ corrections. However
we have not been able to construct a useful measure
with such a behaviour to test this possibility.  

As we remarked above, we expect the $V=\infty$ discontinuity
to be smeared over an $O(1/V)$ transition region
in a finite volume, so that definitions of $\beta_c(V)$ that use, 
for example, the peak of $C(\beta)$ or $\chi_l$ will have 
corrections that are $O(1/V)$. In physical units, we expect 
the correction in the continuum limit to be 
\begin{equation}
{ {T_c(\infty)-T_c(V)} \over {T_c(\infty)} }
\stackrel{V\to\infty}{=}
{ {h^{\prime}} \over {VT_c^3} } + \ldots
\label{C10}
\end{equation}
Inserting $T_c(V)=1/a(\beta_c(V))L_t$ and Taylor expanding 
$a(\beta_c(V))$ around $V=\infty$ this translates, on an $L^3 L_t$ 
lattice, to the conventional finite size extrapolation
\cite{Tsu3} 
\begin{equation}
\beta_c(V) - \beta_c(\infty)
\stackrel{V\to\infty}{=}
h { {L_t^3}\over{L^3} }  
\label{C10b}
\end{equation}
where
\begin{equation}
h 
= 
{ {h^{\prime}}\over{{{d}\over{d\beta}}\ln a(\beta=\beta_c) }}
=
{ {2N^2 h^{\prime}}\over{{{d}\over{d\gamma}}\ln a(\gamma_c) }}.
\label{C10c}
\end{equation}
Thus if the leading finite-$V$ correction is to be finite as
$V\to\infty$ we should find $h\propto N^2$. In
\cite{letterTc} 
we found some evidence that $h$ is constant implying that the
physical finite size correction vanishes as 
$h^{\prime} \propto 1/N^2 $ when $N\to\infty$. 
Note that since $h^\prime$ is a dimensionless physical quantity, 
it receives lattice corrections that are $O(a^2)$. The value
of $h$ on the other hand also depends on how $a$ varies 
with $\beta$, i.e. on the violation of asymptotic scaling.

\subsection{Characterising second order transitions}
\label{subsection_secondorder}

If the deconfining transition is second order, the string tension 
$\sigma(T)$ will vanish smoothly as $T\to T_c^{-}$. The finite-$T$ 
effective potential will have a single minimum for $T<T_c$ with 
a quadratic term that vanishes at $T=T_c$. Thus at this
temperature there is a diverging correlation length. This
will be the inverse mass of the flux loop that winds
around the $t$-torus; the vanishing at $T_c$ of this mass follows
from  eqn(\ref{C4b}) and from the vanishing of the string tension.
For $T>T_c$ the former minimum at the origin becomes a
maximum and a new minimum smoothly develops at a non-zero value
of the order parameter $\langle l_p \rangle$. For SU($N$) 
there would be $N$ degenerate such minima (as discussed in
Section~\ref{subsection_general}); however we know that
only SU(2) is second order so we can specialise to that
case, in which case we have two deconfined vacua characterised
by $\langle l_p \rangle \propto \pm 1$. 

Because the effective potential becomes flat near the 
transition, there will be large fluctuations. These
will however be smooth, in contrast to the large tunnelling
fluctuations of a first order transition. As $T$ increases
above $T_c$ one will begin to have distinct tunnelling
transitions, but these will be between the deconfined vacua.

Just as for first order transitions,  $C(\beta)$ or $\chi_l$ 
diverge at $T_c$ but the $V$ dependence of this divergence
is different and can be used to characterise the transition.
Consider for example the specific heat. We may write it as
\begin{equation}
{1\over{\beta^2}} C(\beta) 
= 
{\partial\over{\partial\beta}} 
\langle \bar{u}_p  \rangle
=
N_p \langle {\bar{u}_p}^2 \rangle 
- 
N_p {\langle \bar{u}_p \rangle }^2
=
N_p \langle (\bar{u}_p - \langle\bar{u}_p\rangle) (\bar{u}_{p} - \langle\bar{u}_p\rangle)  
  \rangle 
\label{C12}
\end{equation}
Since the correlation length diverges as $T\to T_c$, we expect 
that the correlation function of the plaquettes will vary
as $\stackrel{r\to\infty}{\propto}  e^{-m_0(T)r}$ where 
$m_0(T)\to 0$ as $T\to T_c$ and $V\to \infty$. Thus
\begin{equation}
 C(\beta_c(V)) 
\stackrel{V\to \infty}{\propto}
{{V^\zeta}}
\label{C12c}
\end{equation}
where $\zeta \leq 1$ depends on the critical exponents
of the transition through $\zeta=\gamma/\nu d$, using the
standard notation. A similar derivation can be carried out
for the loop susceptibility, $\chi_l$. All this,
as well as the $V$-dependence of $\beta_c(V)$, can be
systematically analysed within the context of finite
size scaling theory
\cite{FSS2}.
%

\subsection{Tunnelling between phases}
\label{subsection_tunnelling}

On an $L^3L_t$ lattice with $L_t\ll L$ we can view  as follows
the tunnelling that occurs between phases that have nearly the 
same free energy. An intermediate step required for tunnelling
is that the $L^3$ periodic volume be split into two roughly equal 
domains each in one of the two phases. Separating these will 
be two domain walls. Since these walls have a finite tension
the most probable configuration will be the one that minimises the 
area, so they will extend right across the short $t$-torus and
across two of the spatial tori. The probability $P_W(T)$ of forming such 
a bubble with two spatial walls of size $l^2=(aL)^2$ is basically
\begin{equation}
P_W(T)
 \propto 
e^{-{{energy \ walls}\over{T}}}
 \propto 
e^{-{{2\sigma_W l^2}\over{T}}}
 \propto 
e^{- 2a^3\sigma_W L^2 L_t}
\label{C15}
\end{equation}
where $\sigma_W$ is the tension of the wall (i.e. energy per unit area). 

We see from the $L$-dependence in eqn(\ref{C15}) that the probability 
of fields with such large domain walls is exponentially suppressed
as the volume grows. This means that the `time' the system
spends in the transition between two phases is very small compared
to the time it remains in one or the other phase: the transitions
are rapid. It also means that the transitions are increasingly
rare as $L$ grows. These rare, large tunnelling fluctuations 
clearly distinguish between first and second order transitions
once we are on a large enough volume. (Such tunnelling
is also of course a feature of spontaneous symmetry breaking in a
finite volume.)

This provides a qualitative description of the transition between 
confined and deconfined phases at $T=T_c$ and between the $N$ 
degenerate deconfined vacua for $T > T_c$. In the latter case,
for $T\gg T_c$ where $g^2(T) \ll 1$, one can use perturbation
theory to evaluate the wall tension. One finds 
\cite{cka_wallN}
for the wall tension between the $k=0$ and $k$'th deconfined
phases, characterised by $\bar{l}_p \sim 1$ and $ e^{i2k\pi/N}$
repectively, 
\begin{equation}
\sigma_W^{0,k}
=
{k(N-k)}{{4\pi^2T^2}\over{3\sqrt{3\lambda(T)}}}
(1+O(g^2N))
\label{C17}
\end{equation}
where $\lambda(T) \equiv g^2(T)N$ is the 't Hooft coupling
that one keeps constant for a smooth large-$N$ limit.

As we observed in 
\cite{letterTc}
the deconfining phase transition appears to become more
strongly first order as $N\uparrow$, and the finite-$V$
corrections appear to become rapidly smaller,
e.g. $h^{\prime}\propto 1/N^2$ in eqn(\ref{C10}). Our more
extensive calculations in this paper will confirm this.
The tunnelling near the phase transition can be 
seen on smaller $V$ as $N\uparrow$, and for a given
$V$ (in physical units) the tunnelling becomes rapidly rarer
as $N\uparrow$. We interpret this as being due to the 
confining/deconfining interface tension, $\sigma_{c,d}$,
growing with $N$. A reasonable guess is that for
$T\simeq T_c$ the tunnelling between the $k=0$
and $k=N/2$ deconfined phases (say $N$ is even), will   
effectively pass through a sheet of confined phase so
that 
\begin{equation}
\sigma_{c,d}
\propto
\frac{1}{2} \sigma_W^{k=0,k=N/2}
\propto 
N^2
\label{C19}
\end{equation}
using eqn(\ref{C17}). From eqn(\ref{C15}) we see that
this means that the tunnelling has a probability
\begin{equation}
P_{c,d}(L,T)
\propto
e^{-{{cN^2 L^2}\over{T}}}
\label{C21}
\end{equation}
so that to see tunnelling equally easily we need to
reduce the volume $L\propto 1/N$ as we increase $N$.
This roughly describes what we shall see in our calculations.

The assumptions of our above argument can, in fact, be weakened
considerably. All we need is that
$\sigma_W^{k=0,k=N/2} \sim O(N^2)$ for  $T\simeq T_c$,
and then the fact that  $\sigma_W^{k=0,k=N/2} \leq 2\sigma_{c,d}$
obtains for us eqn(\ref{C19}) as a lower bound on the $N$-dependence
of the domain wall tension.
We remark that the growth of domain wall surface tensions as 
a power of $N$ appears to be a common phenomenon 
(see Section~\ref{section_conjectures}) so it is no surprise
that it also occurs in the case of the deconfining
phase transition.

\section{Results}
\label{section_results}

\subsection{Setting the scale}
\label{subsection_scale}

To set the scale of the lattice spacing in physical units 
we use the $T=0$ string tension, $\sigma$. We extract $\sigma$
from the mass of the lightest flux loop that winds around
the spatial torus, which can be obtained from correlations 
of smeared Polyakov loops as described, for example, in
\cite{blmt-kstring}.
We generally use spatial tori of length  
$l\equiv aL \geq 3/\surd\sigma$, since earlier calculations
\cite{blmt-kstring}
suggest that finite volume corrections are then accurately
given by just the leading universal string correction, i.e.
\begin{equation}
am_p(L)
=
a^2\sigma L - {{\pi}\over{3 L}}
\label{D2}
\end{equation}
assuming the confining string is a simple bosonic string
as suggested by 
\cite{blmt-kstring,LWstring}.

We have performed string tension calculations for all
the SU($N$) groups discussed in this paper. The
calculations are taken from a study of improved gluonic
lattice operators 
\cite{newK}
and we use a type of blocking/smearing that is much
more effective than in earlier calculations. This leads
to significant changes in earlier estimates of the
continuum value of $T_c/\surd\sigma$ for both SU(2) and
SU(3).

Our calculated values of $a\surd\sigma$ are listed in
Table~\ref{table_su23sigma} and Table~\ref{table_su468sigma}.
Once we determine the critical coupling $\beta_c(\infty)$
for a given value of $L_t$, we interpolate our values of 
$a\surd\sigma$ to that value of $\beta$ and this provides
our lattice estimate of $T_c$
\begin{equation}
{{aT_c} \over {a\surd\sigma}}
=
{{T_c} \over {\surd\sigma}}
=
{1 \over {L_t a(\beta_c)\surd\sigma}}
\ \ \ \ ; \ \ \ 
a(\beta_c) = {1\over{L_tT_c}}.
\label{D4}
\end{equation}
Repeating the calculation for various $L_t$, we
can attempt to extrapolate the results to the continuum
limit using a leading $O(a)$ lattice correction
\begin{equation}
{{T_c(a)} \over {\surd\sigma(a)}}
=
{{T_c(a=0)} \over {\surd\sigma(a=0)}}
+
c a^2\sigma
\label{D6}
\end{equation}
where $c$ is some unknown quantity that is fitted.

\subsection{Locating the phase transition}
\label{subsection_locating}

To locate a first order transition we search for values
of $\beta$ at which there is tunnelling between the
confined and deconfined phases. (See e.g.~Fig.1 and Fig.6 of 
\cite{letterTc}.)
This is easiest to do on smaller volumes where the tunnelling 
is sufficiently frequent that it can be seen in modest Monte 
Carlo runs. (Although if the volume is too small the 
transition will be completely washed out.) One can
then use the $\beta_c(V)$ extracted from a smaller volume
as a starting point for calculations at larger $V$. 

\begin{figure}[htb]
\begin{center}
\includegraphics[angle=-90,width=\figsize]{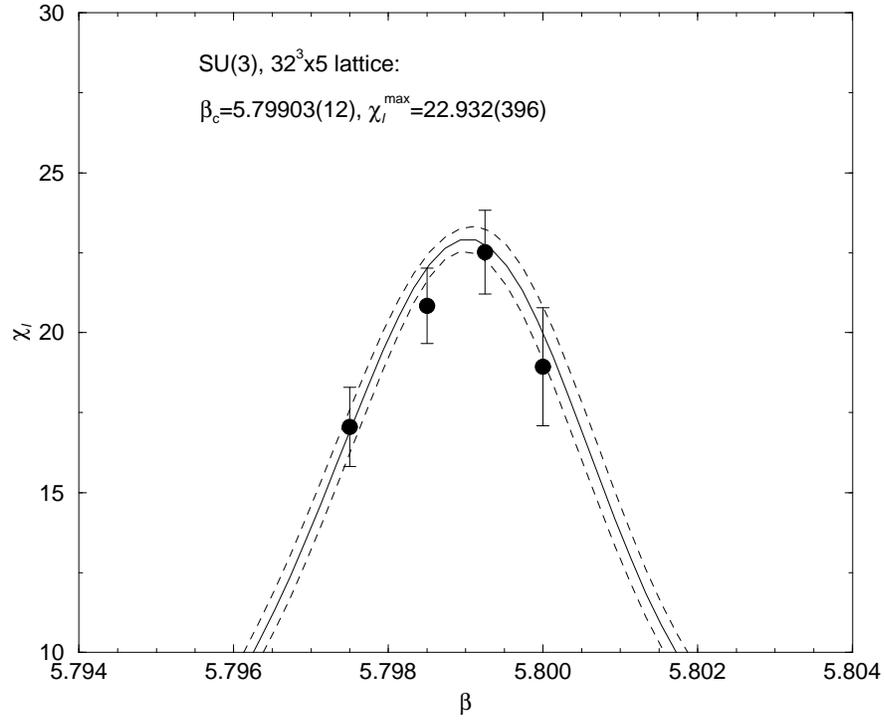} 
\end{center}
\caption[]{Reweighted susceptibility $\chi_l$ as a function of $\beta$
for SU(3) on a $L=32$ lattice at $L_t=5$. The location of the maximum,
$\beta_c$, and the value of the maximum, $\chi_l^\text{max}$, are given
as well.}
\label{fig:susc_su3_l32_t5}
\end{figure}

We typically carry out $O(200K)$ sweeps at each $\beta$
and have 2 to 5 useful values of $\beta$. They will be
close enough together that the distributions of $\bar{u}_p$ 
have a strong overlap, and we then use standard reweighting
techniques 
\cite{reweight}
to produce $\chi_l(\beta)$ and $C(\beta)$ as continuous functions 
of $\beta$ and to obtain an accurate value for the position of
the maximum, $\beta^C_c(V)$ or  $\beta^\chi_c(V)$. An example is
shown in Fig.\ref{fig:susc_su3_l32_t5}. We then repeat
the calculation for various volumes, $L^3$, at a fixed value
of $L_t$ and perform a finite size study (see below) to obtain 
$\beta_c(\infty)$, using eqn(\ref{C10b}), the latent heat, 
using eqn(\ref{C6}), and so on. We then proceed to the
lattice and continuum values of $T_c/\surd\sigma$ as
described in Section~\ref{subsection_scale}.

In practice we carry out the finite volume study at one
value of $L_t$ for each value of $N$ and then use the
parameter $h$ obtained from a fit of the form in 
eqn(\ref{C10b}) at the other values of $L_t$ where we
perform calculations typically on just one volume.
This is the procedure used in past SU(3) calculations, e.g.
\cite{Tsu3},
but is dangerous because the corrections to $h$ in
eqn(\ref{C10b}) are not merely $O(a^2)$, as they
would be for $h^{\prime}$ in eqn(\ref{C10}), but
include the violations of asymptotic scaling
in the $d\ln a / d\beta $ factor. This is a
weakness of this calculation, which can only be remedied 
by a study on a much larger scale than ours. 

For the case of SU(2), where the transition is second order,
it turns out that the specific
heat $C(\beta)$  has no visible peak as we vary $\beta$ and 
its value does not grow with $V$. This anomalous behaviour
arises because the plaquette has almost no overlap onto
any correlation length that diverges. This interesting
behaviour will be discussed in more detail in 
\cite{laterTc}.
Here we merely note that one can instead focus upon
the Polyakov loop susceptibility $\chi_l$, as is usually done, 
and that exhibits all the characteristics of a second order
phase transition.

\subsection{Finite size study at $L_t=5$}
\label{subsection_finitesize}

We perform our finite-$V$ studies at $L_t=5$ for all the values 
of $N$ that we investigate. For a quick global view of what is going on we
plot in Fig.2 the value of $\ln \chi_l$ (extracted at its peak
value) against $\ln V$ for all our values of $N$. We see that
for $N\geq 3$ the values fall approximately on a straight line of 
slope unity. This is just the slope appropriate to a first order 
transition (see eqns(\ref{C6b},\ref{C8b})) and is in contrast
to the SU(2) line that is flatter (with a slope of $\sim 0.63$)
indicating that the transition there must be second order.

To see the corrections, a plot of the normalised susceptibility maxima
$\chi_l^\text{max}/V$ versus $1/V$ is
more useful, and this we show in
Fig.\ref{fig:chi_max_scaling_su2_su3_t5} for SU(2) and SU(3). We see that for
SU(2)  $\chi_l^\text{max}/V \to 0$ as $V\to\infty$ indicating it
is second order. (Our largest SU(2) lattice is $L=64$.)
For SU(3) the behaviour at smaller $V$ is similar to that
of SU(2), indicating a growing correlation length,  but eventually 
it turns upwards, showing that it is indeed dominated by a 
first-order discontinuity. So SU(3) is `weakly' first order.
For $N\geq 4$ (Fig.\ref{fig:chi_max_scaling_su4_su6_t5}) the picture is much less ambiguous: there is the 
expected correction that is linear in $1/V$ with a flattening
only at small $V$ where presumably the correlation length is
being felt. As $N\uparrow$ the values of $V$ where the flattening
occurs is at ever smaller $V$ indicating that the strength of the
first order transition is growing (in the sense that the
discontinuity dominates over the contribution of the longest
correlation length at ever smaller $V$).

In Fig.\ref{fig:c_max_scaling_su3_su4_su6_t5} we perform similar plots for the normalised specific heat maxima
$C_\text{max}/V$. We see a much nicer linear behaviour that
is easy to extrapolate. The reason for this is that,
unlike $\chi_l$, $C(\beta)$ is not sensitive to the
lightest correlation length (which is proportional to
$\sigma(T)$). Indeed we do not show the SU(2) curve,
because we cannot identify a maximum -- and if we did
its value would be independent of $V$ (in our range of $V$).

We can use the maximum of $\chi_l$ to define $\beta_c(V)$
and from a fit to eqn(\ref{C10b}) we can extract $h$, the coefficient
of the $1/V$ correction. This is tabulated in Table~\ref{table_h}.
We recall that the usually quoted SU(3) value is $h\leq 0.1$
\cite{Tsu3}.
We find it impossible to extract a reliable value of $h$
from our values of $\beta_c(V)$ at $L_t=5$ (although we
have some calculations for $L_t=4$ that agree precisely with
earlier calculations) and we put that down to the weakness
of the SU(3) transition -- or perhaps a large statistical
fluctuation in our calculations. In any case, it is already
clear from  Table~\ref{table_h} that we do not have 
$h \propto N^2$ as one would need for a finite $h^{\prime}$
in the $N=\infty$ limit. (See eqns(\ref{C10},\ref{C10b},\ref{C10c}).)
Rather it looks as though $h^{\prime} \propto 1/N^2$.

\subsection{Latent heat}
\label{subsection_latentheat}

We see from eqn(\ref{C6}) that we can obtain $L_h$, the
latent heat per plaquette, from the $V=\infty$ extrapolation
of $C_\text{max}/V$. Doing so for our $L_t=5$ calculations,
we obtain the value of $L_h$ for each $N$ and we plot the results in 
Fig.\ref{fig:latent_heat_vs_N}.

The first thing we observe is that  $L_h$ grows with $N$.
Moreover the growth is consistent with being due to
a $O(1/N^2)$ correction, as shown by the dashed line in the
plot. It is clearly no surprise that the latent heat vanishes
for $2<N<3$, as exemplified by our fit, so that the $N=2$
transition is second order.

The lattice latent heat is $O(a^4)$. Can a variation of
$a$ with $N$ account for the variation we see in
Fig.\ref{fig:latent_heat_vs_N}? 
If we express $a$ in units of $T_c$ then the answer is of course 
no: by definition $a=1/L_tT_c =1/5T_c$ is fixed for all $N$.
If instead we use $\sigma$ to set the scale, then $a^4$
does indeed vary as we go from $N=3$ to $N=6$, but only
by about $\sim 30\%$ which is a very small part of the
observed change. Thus we conclude that in physical units
the latent heat increases strongly as we go from $N=3$
to larger $N$, indicating that the first order transition
strengthens as $N\uparrow$.   

Finally we remark that if the latent heat goes to a
non-zero constant as $N\to\infty$, this means
that it is proportional to the gluon condensate 
(with the 't Hooft running coupling) 
\cite{Gcondensate}
so that it is $O(N^2)$.

\subsection{$T_c/\surd\sigma$ for all $N$}
\label{subsection_Tc}

We list our values of $\beta_c(V=\infty)$ in 
Table~\ref{table_su23Tclatt} and Table~\ref{table_su468Tclatt}.
The SU(2) values include older calculations
\cite{Tsu2}
as do the SU(3) ones
\cite{Tsu3,Tsu3_jap}.
The (interpolated) string tensions are new and more accurate
than those used previously. For our SU(4) and  SU(6)
calculations at $L_t=6,8$ we use the $h$ extracted at $L_t=5$.
However we double the error so as to (hopefully) cover any
change of $h$ with $L_t$. For SU(8) we also extract $h$ for $L_t=5$, 
but because the error on it is already very large, we feel no
need to increase it at larger $L_t$.

We use eqn(\ref{D4}) to obtain the values of  $T_c/\surd\sigma$
shown, and then use  eqn(\ref{D6}) to obtain the continuum
extrapolation, as tabulated in Table~\ref{table_Tcsigma_cont}.
(In the SU(4) case where the best $\chi^2$ is very poor, we 
use an error that covers the extrapolated values obtained
using any two $\beta$ values out of the three.) 

Finally we take our various continuum values of  $T_c/\surd\sigma$
and plot them against $1/N^2$ in Fig.~\ref{fig:tc_sigma_vs_N}. We see
that the values are all well described by a leading
large-$N$ correction of the expected 
\cite{largeN}
functional form:
\begin{equation}
{{T_c} \over {\surd\sigma}}
= 
0.596(4) + {{0.453(30)}\over{N^2}}.
\label{D8}
\end{equation}
The coefficient is modest, indicating that for this physics at
least, all values of $N$ are close to $N=\infty$, despite
the fact that SU(2) is second order while larger groups
are first order. Only the latent heat appears to have
large corrections in $N$.

\subsection{$V$-dependence as $N\uparrow$}
\label{subsection_volume}

The values of $h$ in Table~\ref{table_h} strengthen
our earlier observation
\cite{letterTc}
that  $h$ is roughly constant indicating that the
finite volume corrections disappear as $\propto 1/N^2$.

It is also clear from e.g.~Fig.\ref{fig:chi_max_scaling_su2_su3_t5}
and \ref{fig:chi_max_scaling_su4_su6_t5} that the first order
character of the transition becomes clearer on ever smaller
volumes as $V\uparrow$. We have interpreted this as 
being related to the domain wall surface tension growing 
with $N$, perhaps as in eqns(\ref{C17},\ref{C19}).
To analyse this further we have roughly calculated the
number of sweeps that the lattice spends in each of the confined
and deconfined phases when it is very close to $\beta_c(V)$.
We call this number the persistence time, $\tau_p$.
From the statistical errors on our $T=0$ string tension
calculations, we infer that the efficiency of our
Monte Carlo in decorrelating physical fluctuations
does not vary much with $N$, so that comparing  
$\tau_p$ across $N$ is physically meaningful. 
We plot  $\tau_p$ for various volumes and various $N$ in Fig.\ref{fig:persistence_time_sun},
against $L^2$. We see that it is consistent with being
$\tau_p \propto \exp{-(N^{\alpha} L)^2}$  with $\alpha$ 
unity or greater. In fact the $L$ dependence is not really well
determined: a plot against $L^3$ looks equally good.
If we simply take the largest $L$ where we have roughly
equal $\tau_p$ for $N=3,4,6$ i.e.~$L=32,20,12$ respectively,
then a trend to be approaching a function of $N\times L$ for
larger $N$ is certainly consistent. However this analysis
has to be regarded as very preliminary at this stage,
and only indicative of what one might eventually achieve
\cite{laterTc}.
%

\section{Some conjectures about $N=\infty$}
\label{section_conjectures}

\subsection{Reduced models}
\label{subsection_reduced}

On a given spatial volume, $V=L^3$, at fixed $L_t$ and at values 
of $\beta$ close to the corresponding phase transition, we
have found the tunnelling to be rapidly suppressed as $N\uparrow$.
Roughly speaking if one wants to keep the tunnelling rate
constant one needs to decrease the volume as $L\propto 1/N^{\alpha}$,
where $\alpha$ appears to be at least unity. This is roughly 
consistent with our conjecture that the surface tension 
of the confining/deconfining domain wall varies as 
$\sigma_{c,d} \propto N^2$, and in any case that this provides 
a lower bound on its variation with $N$. We also saw that
the shift of $aT_c(V)$ from $aT_c(\infty)$ appears to
be $O(1/N^2)$, suggesting that as $N\to\infty$ one can
locate the transition on volumes $V\to 0$.

As we pointed out in
\cite{letterTc} 
this cannot really be the case, because once 
$L < 1/a(\beta)T_c \simeq L_t$
the spatial version of the center symmetry in eqn(\ref{C4c})
will become spontaneously broken. For example if we are 
considering $L_x < L_t$ 
then the relevant symmetry is as in eqn(\ref{C4c})
but with $U_x$ replacing $U_0$, $n_x$ replacing $n_t$,
and $n_s$ counting the points in the $(y,z,t)$ 3-slice.
This breaking occurs for the same reason that the symmetry is
broken in the temporal direction for $L_t < 1/a(\beta)T_c$;
after all the label of the Euclidean direction does not matter.
(For finite $N$ this breaking will itself be smoothened by
finite volume effects; however as $N\to\infty$ the observed
loss of finite-$V$ corrections suggests it will become precise,
even on a finite volume.) For this reason we conjectured in
\cite{letterTc} 
that as $N\to\infty$ the deconfining transition will occur at 
exactly the same $\beta_c$ for all volumes $L > L_t = 1/a(\beta)T_c$,
i.e.~the deconfining physics is independent of the volume down to a 
critical volume, and we speculated upon the relation of this
to Witten's idea of an $N=\infty$ Master Field that is
necessarily translation invariant
\cite{Master}. 
We also note the relation of this to recent conjectures 
\cite{neubergerEK}
about how the Eguchi-Kawai reduction  
\cite{EK} 
translates to the continuum.

Of course, as $N\to\infty$ the transitions on this minimal $V$
will presumably become rapidly rarer and the transition will
become ever harder to locate. 
Here we speculate on a way of reducing $V$ below this minimal value.
This is based on the observation that if we impose an $\{x,t\}$ 
twist for $T>T_c$, it induces a corresponding $Z_N$ domain
wall separating two of the $Z_N$ deconfined phases. That is
to say, it disorders the $Z_N$ symmetry breaking. Indeed this
domain wall can be interpreted 
\cite{vortexwall}
as a 't Hooft dual vortex in its `Higgs' phase. If $L_x$
becomes small there is no room for the two phases and the wall, 
and we expect that the corresponding symmetry breaking 
is suppressed. Thus we conjecture that on an $L^3 L_t$
volume at $\beta_c$, if we reduce $L < L_t$ but introduce
a (maximal) twist in   $\{x,y\}$,  $\{x,z\}$ and $\{y,z\}$ we
suppress the symmetry breaking and can continue to
reduce $L$, perhaps as $1/N$, while maintaining the physics of the
deconfining transition. As $N\to\infty$ we can therefore
study $T_c$ on an infinitesimal volume: in a sense on
a single spatial point. And perhaps other physical quantities
can also be so studied.
This reduction connects naturally to the twisted Eguchi Kawai 
reduced model
\cite{EK,TEK}. 
%

\subsection{Master field(s)}
\label{subsection_master}

The factorisation of gauge invariant operators at large-$N$,
and the fact that fluctuations vanish in leading order,
led to an old speculation 
\cite{Master} 
that in the $N=\infty$ limit there is only a single Master field
in the path integral (which is necessarily translation invariant 
for gauge invariant quantities). It has long been realised that 
this picture is deceptively simple (e.g.
\cite{largeN2}).

In the case of interest here, we have seen that the fluctuations 
near $T=T_c$ between the confined and deconfined phases, 
are suppressed exponentially in $N^2$ (at least) on a volume
that is fixed in physical units as $N\uparrow$. If one wishes
to talk of a `Master Field' one has to talk of at least 2:
one confined and one deconfined. (For the D=3+1 Euclidean
system there are presumably $N\to\infty$ deconfined Master
Fields, separated by infinite barriers in that limit.)

We note that as $T$ deviates from $T_c$ the difference in free 
energy density between the confined and deconfined vacua will
be $O(N^2)$ and for large enough bubbles will outweigh the 
suppression of the domain walls (if that is also $O(N^2)$ as 
we have suggested). However the system has
to pass through small bubbles to reach these large bubbles,
and the probability of these will continue to be dominated
by the exponential suppression of the surface area of the
bubble. That is to say, one can imagine a scenario at $N=\infty$
in which there is no transition between confining and deconfining
phases at any value of $T$ -- a kind of infinite
super-cooling/heating.

This picture of different phases separated by infinite barriers
at $N=\infty$ is also characteristic of the lattice bulk transition
\cite{laterTc,blmt-glue}
and of the multiple vacua inferred from the $\theta$-dependence of
the SU($N$) gauge theory
\cite{witten_vacN,shifman_vacN}.
All this suggests that the gauge theory possesses an infinite
number of `Master Fields' that mutually decouple precisely
at $N=\infty$ but whose simultaneous existence ensures a rich 
physics of that limiting theory.

\section{Conclusions}
\label{section_conclusions}

In this paper we have obtained some properties of the deconfining 
transition in SU($N$) gauge theories. We have done so for 
enough values of $N$ that we feel confident in our control of the 
$N\to\infty$ limit. At the same time, our dedicated string tension
calculations allow us to improve significantly upon older
continuum extrapolations for SU(2) and SU(3).

We find the phase transition is first order for $N\geq 3$,
and that the latent heat (of the $D=3+1$ Euclidean system) grows
with $N$, as shown in Fig.\ref{fig:latent_heat_vs_N}, showing that the transition
becomes more strongly first order as $N\uparrow$. Indeed it
is clear from Fig.\ref{fig:latent_heat_vs_N} that the SU(3) transition is quite
weakly first-order, and that a naive interpolation of
the latent heat as a function of $N$ predicts that the
transition ceases to be first order somewhere between $N=2$
and $N=3$. 

We also find that as  $N$ grows the phase transition becomes
clearly defined on ever smaller volumes (see e.g.~Fig.\ref{fig:chi_max_scaling_su2_su3_t5}-\ref{fig:c_max_scaling_su3_su4_su6_t5}). 
We argue that this is due to the interface tension between the 
confined and deconfined phases growing at least as rapidly as $N^2$. 
At the same time we find that the finite-$V$ corrections to
$T_c$ appear to be $O(1/N^2)$ so that at $N=\infty$ $T_c$ can
be calculated on small volumes. We speculated in 
Section~\ref{section_conjectures} how one might be able to take
the limit $V\to 0$ in a way that is reminiscent of the twisted
Eguchi-Kawai model. We also speculated on how this might fit in
with the old idea that $N=\infty$ is dominated by a Master Field
that is translation invariant for gauge invariant observables.
We plan to test the idea in Section~\ref{subsection_reduced},
not only for $T_c$ but for other physical quantities as well.

The values of $T_c/\surd\sigma$ that we obtain in the continuum
limit show a modest $N$-dependence, displayed in Fig.~\ref{fig:tc_sigma_vs_N}.
This dependence can be well fitted with a leading $O(1/N^2)$ 
correction all the way down to SU(2), as in eqn(\ref{D8}).
Such modest corrections are in fact characteristic of the $N$
dependence of many physical quantities
\cite{blmt-glue,PisaQ}.

As we have stressed here and in
\cite{letterTc}
there are many ways in which our calculations need to  be improved.
Perhaps their main weakness lies in the control 
of finite-$V$ corrections. Because such studies are expensive we
have followed the usual strategy of performing a finite-$V$
study only at the coarsest value of the lattice spacing $a$ and then 
using that to estimate finite-$V$ corrections at other values of $a$.
That this is likely to be dangerous is indicated by the nature of
the theoretical expression, eqn(\ref{C10c}). Another weakness
concerns our latent heat calculation which, being coupled to
the finite-$V$ study, has again only been performed for the
coarsest value of $a$. Here in fact we can do better, using a
different technique, and in  
\cite{laterTc}
we will do so. In that paper we will also present calculations
of the correlation lengths as $T\to T_c$, obtained separately
for the confined and deconfined phases. This will provide
another measure of the strength of the transition. In 
addition we will present some results for the physical latent
heat of the gauge theory (which is different from the `Euclidean'
latent heat that we discuss in this paper), for the way topological
fluctuations change across $T=T_c$, and we will present the results
discussed in this paper in greater detail.

\section*{Acknowledgments}

We appreciate discussions with C.P.~Korthals Altes and H.~Neuberger.
Our lattice calculations were carried out on the JIF/PPARC funded 
Astrophysics Beowulf cluster in Oxford Physics, on PPARC 
and EPSRC funded Alpha Compaq workstations in Oxford Theoretical 
Physics,  on a desktop funded by All Souls College, and on the 
APE in Swansea Physics funded by PPARC under grant PPA/G/S/1999/00026.
UW has been supported by a PPARC fellowship, and BL by a
EU Marie Sklodowska-Curie postdoctoral fellowship.


\newpage

\begin{table}
\begin{center}
\begin{tabular}{|c|c|c||c|c|c|}\hline
\multicolumn{3}{|c||}{SU(2)} & \multicolumn{3}{c|}{SU(3)} \\ \hline
$\beta$ & $L$ & $a\surd\sigma$ & $\beta$ & $L$ & $a\surd\sigma$  \\ \hline
1.8800  & 4  & 0.773(16)  & 5.6925 & 8  & 0.3970(19)  \\
2.1768  & 8  & 0.5149(77) & 5.6993 & 8  & 0.3933(16)  \\
2.2986  & 10 & 0.3667(18) & 5.7995 & 10 & 0.3143(14)  \\
2.3726  & 12 & 0.2879(10) & 5.8945 & 12 & 0.2607(11)  \\
2.4265  & 16 & 0.2388(9)  & 6.0625 & 16 & 0.19466(73)  \\
2.5115  & 20 & 0.17680(76)& 6.3380 & 24 & 0.12946(61)  \\ \hline
\end{tabular}
\caption{\label{table_su23sigma}
SU(2) and SU(3) string tensions calculated at the indicated values of 
$\beta$ from confining strings of length $aL$ that wind around the
spatial torus.}
\end{center}
\end{table}

\begin{table}
\begin{center}
\begin{tabular}{|c|c|c||c|c|c||c|c|c|}\hline
\multicolumn{3}{|c||}{SU(4)} & \multicolumn{3}{c||}{SU(6)} & 
\multicolumn{3}{c|}{SU(8)} \\ \hline
$\beta$ & $L$ & $a\surd\sigma$ & $\beta$ & $L$ & $a\surd\sigma$ 
& $\beta$ & $L$ & $a\surd\sigma$  \\ \hline
10.637 & 10 & 0.3254(11)  & 24.500 & 10 & 0.3420(19)  & 43.85 & 8  &
0.3646(25) \\
10.789 & 12 & 0.27015(86) & 24.845 & 12 & 0.2801(13)  & 44.00 & 10 &
0.3406(20) \\
11.085 & 16 & 0.19871(85) & 25.452 & 16 & 0.20992(82) & 44.35 & 10 &
0.2991(20) \\
       &    &             &        &    &             & 44.85 & 12 &
0.2596(24) \\ \hline
\end{tabular}
\caption{\label{table_su468sigma}
SU(4), SU(6) and SU(8) string tensions calculated at the indicated values of 
$\beta$ from confining strings of length $aL$ that wind around the
spatial torus.}
\end{center}
\end{table}

\begin{table}
\begin{center}
\begin{tabular}{|c|c|cc|}\hline
   &  $h$  & $L\in$ & $\chi^2/n_{df}$\\ \hline
SU(4) & 0.090(17) & [12,20] & 0.49  \\ 
SU(6) & 0.112(19) & [8,14]  & 0.04  \\ 
SU(8) & 0.063(70) & [6,8]   & --     \\ \hline
\end{tabular}
\caption{\label{table_h}
The coefficient $h$ of the $1/V$ correction to
$\beta^{\chi}_c(V)$ for the groups shown, obtained
from fits (best $\chi^2$ shown) to the range of
volumes shown.}
\end{center}
\end{table}

\begin{table}
\begin{center}
\begin{tabular}{|c|c|c||c|c|c|}\hline
\multicolumn{3}{|c||}{SU(2)} & \multicolumn{3}{c|}{SU(3)} \\ \hline
$L_t$  & $\beta_c$ & $T_c/\surd\sigma$ & $L_t$ & $\beta_c$ & $T_c/\surd\sigma$ \\ \hline
 2 & 1.8800(30)         & 0.647(15)   & --  & --  & --  \\
 3 & 2.1768(30)         & 0.6474(111) & 4  & 5.69236(15) & 0.6296(30) \\
 4 & 2.2986(6)          & 0.6818(35)  & 5  & 5.8000(5)$^\star$   & 0.6370(29) \\
 5 & 2.37136(54)$^\star$ & 0.6918(29)  & 6  & 5.8941(12)  & 0.6388(32) \\
 6 & 2.4271(17)$^\star$  & 0.6994(50)  & 8  & 6.0625(18)  & 0.6421(31) \\
 8 & 2.5090(6)$^\star$   & 0.7008(34)  & 12 & 6.3385(55)  & 0.6442(51) \\ \hline
\end{tabular}
\caption{\label{table_su23Tclatt} Critical values of $\beta$,
extrapolated to $V=\infty$, for SU(2) and SU(3) for the values
of $L_t$ shown; with the corresponding values of the deconfining
temperature in units of the string tension. Starred values are
new calculations of $\beta_c$.}
\end{center}
\end{table}

\begin{table}
\begin{center}
\begin{tabular}{|c||c|c||c|c||c|c|} \hline
\multicolumn{1}{|c||}{} & \multicolumn{2}{c||}{SU(4)} & 
\multicolumn{2}{c||}{SU(6)} & \multicolumn{2}{c|}{SU(8)} \\ \hline
$L_t$  & $\beta_c$ & $T_c/\surd\sigma$ & $\beta_c$ &
$T_c/\surd\sigma$ & $\beta_c$ & $T_c/\surd\sigma$ \\ \hline
 5 & 10.63709(72) & 0.6146(23) & 24.5139(24) & 0.5894(36) & 43.978(22)
& 0.5814(68) \\
 6 & 10.7893(23)  & 0.6171(26) & 24.8458(33) & 0.5952(32) & 44.496(3) 
& 0.5807(71) \\
 8 & 11.0848(23)  & 0.6289(31) & 25.4712(62) & 0.6009(29) & 45.606(35)
& 0.5964(154) \\ \hline
\end{tabular}
\caption{\label{table_su468Tclatt} Critical values of $\beta$,
extrapolated to $V=\infty$, for SU(4), SU(6) and SU(8) for the values
of $L_t$ shown; with the corresponding values of the deconfining
temperature in units of the string tension.}
\end{center}
\end{table}

\begin{table}
\begin{center}
\begin{tabular}{|c|c|c|c|}\hline
  & $lim_{a\to 0} T_c/\surd\sigma$  & $L_t\in$ & $\chi^2/n_{df}$ \\ \hline
SU(2) & 0.7091(36) & [3,8]  & 0.28  \\ 
SU(3) & 0.6462(30) & [4,12] & 0.05 \\ 
SU(4) & 0.634(12)  & [5,8]  & 2.05 \\ 
SU(6) & 0.6078(52) & [5,8]  & 0.00  \\ 
SU(8) & 0.594(20)  & [5,8]  & 0.53  \\ \hline
SU($\infty$)  & 0.5960(41)  & -- & 0.29 \\ \hline
\end{tabular}
\caption{\label{table_Tcsigma_cont}
Continuum extrapolations of $T_c/\surd\sigma$ with the range of
$L_t$ values used and the $\chi^2$ per degree of freedom of
the fit. Also shown is the result of the $N\to\infty$ 
extrapolation of these values, as described in the text.} 
\end{center}
\end{table}

\clearpage

\begin{figure}[htb]
\begin{center}
\includegraphics[angle=-90,width=\figsize]{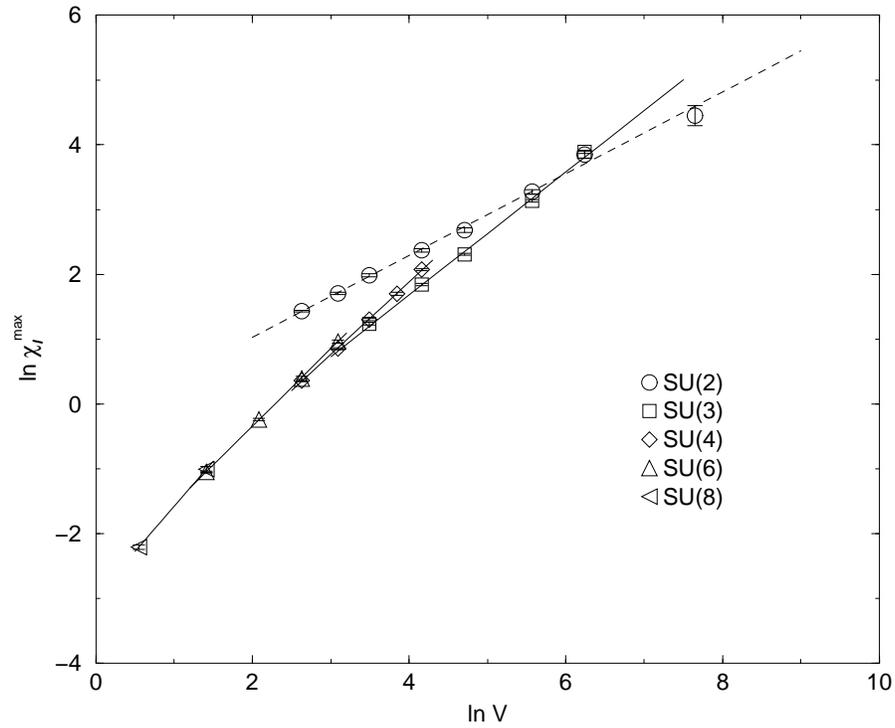} 
\end{center}
\caption[]{The value of $\ln \chi_l^\text{max}$ plotted versus $\ln V$
for all our values of $N$ at $L_t=5$. The full lines are best fits for
$N=3,4,6,8$ with a slope close to unity while the dashed line is the
best fit for $N=2$ with a slope of $\sim 0.63$.}
\label{fig:crit_exp_nt5_sun}
\end{figure}

\begin{figure}[htb]
\begin{center}
\includegraphics[angle=-90,width=\figsize]{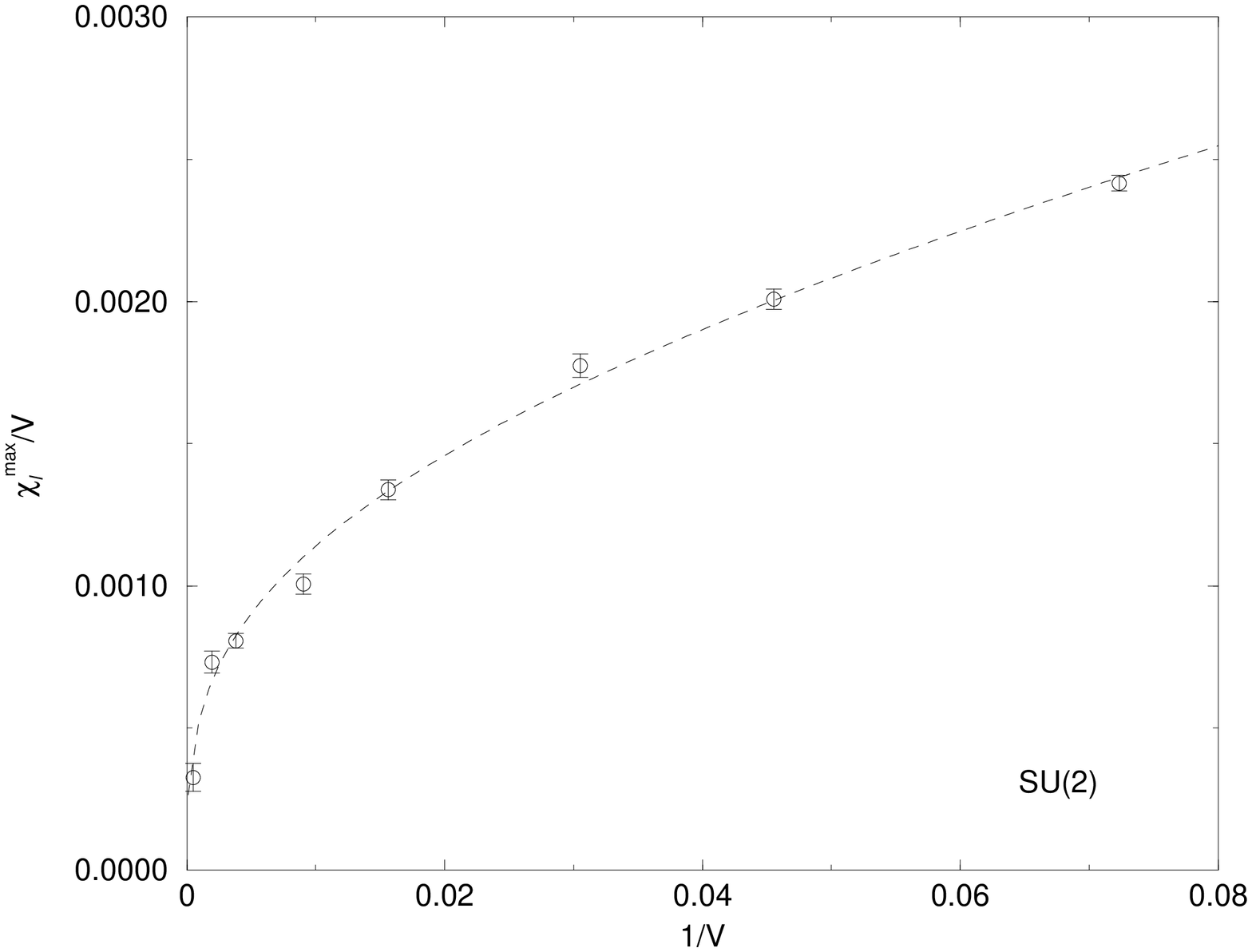}\\
\includegraphics[angle=-90,width=\figsize]{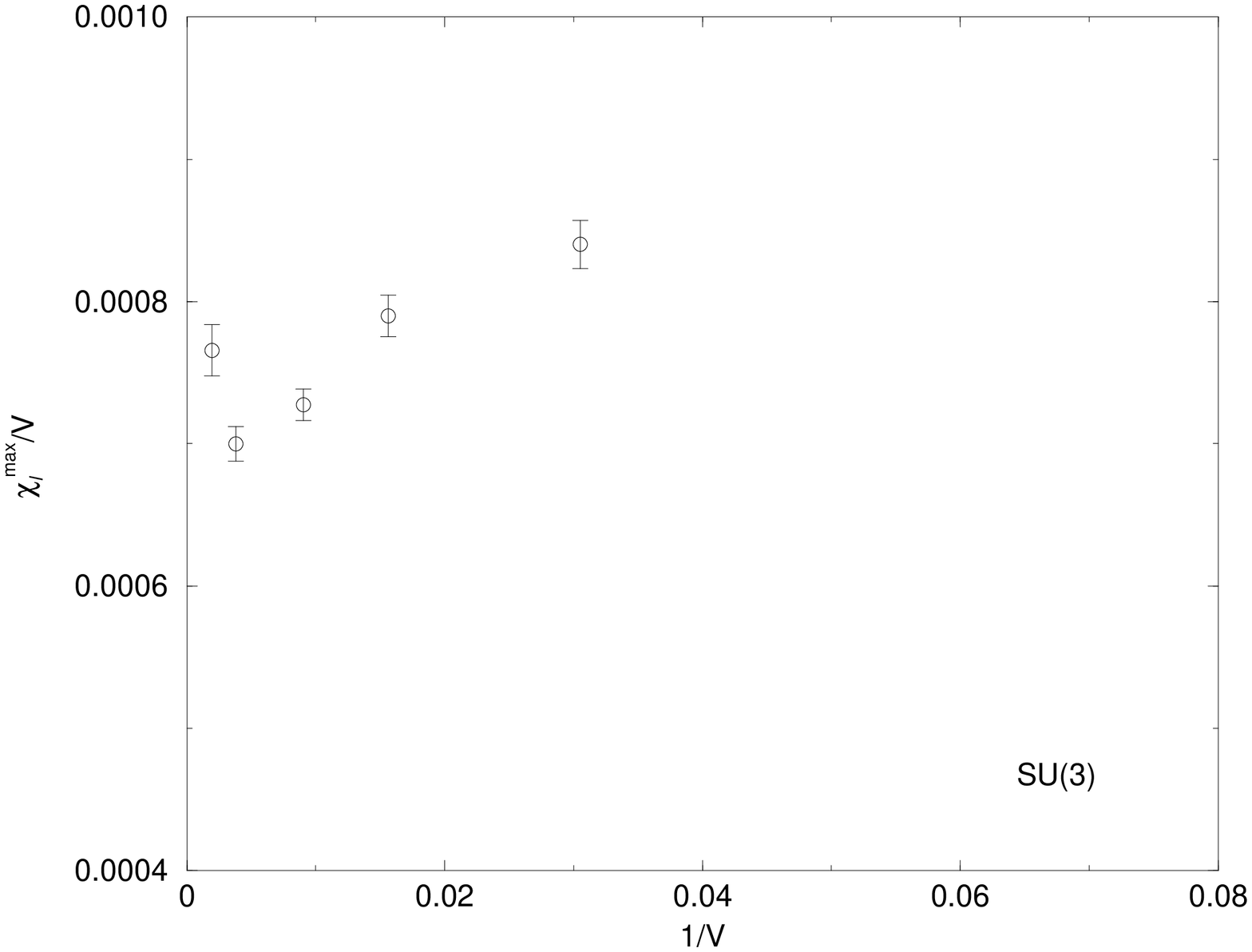}\\
\end{center}
\caption[]{The normalised susceptibility maxima
$\chi_l^\text{max}/V$ plotted against $1/V$ for $N=2,3$ at
$L_t=5$. The dashed line is a best fit with the leading scaling
behaviour extracted from Fig.\ref{fig:crit_exp_nt5_sun} and a leading
$O(1/V)$ correction.}
\label{fig:chi_max_scaling_su2_su3_t5}
\end{figure}

\begin{figure}[htb]
\begin{center}
\includegraphics[angle=-90,width=\figsize]{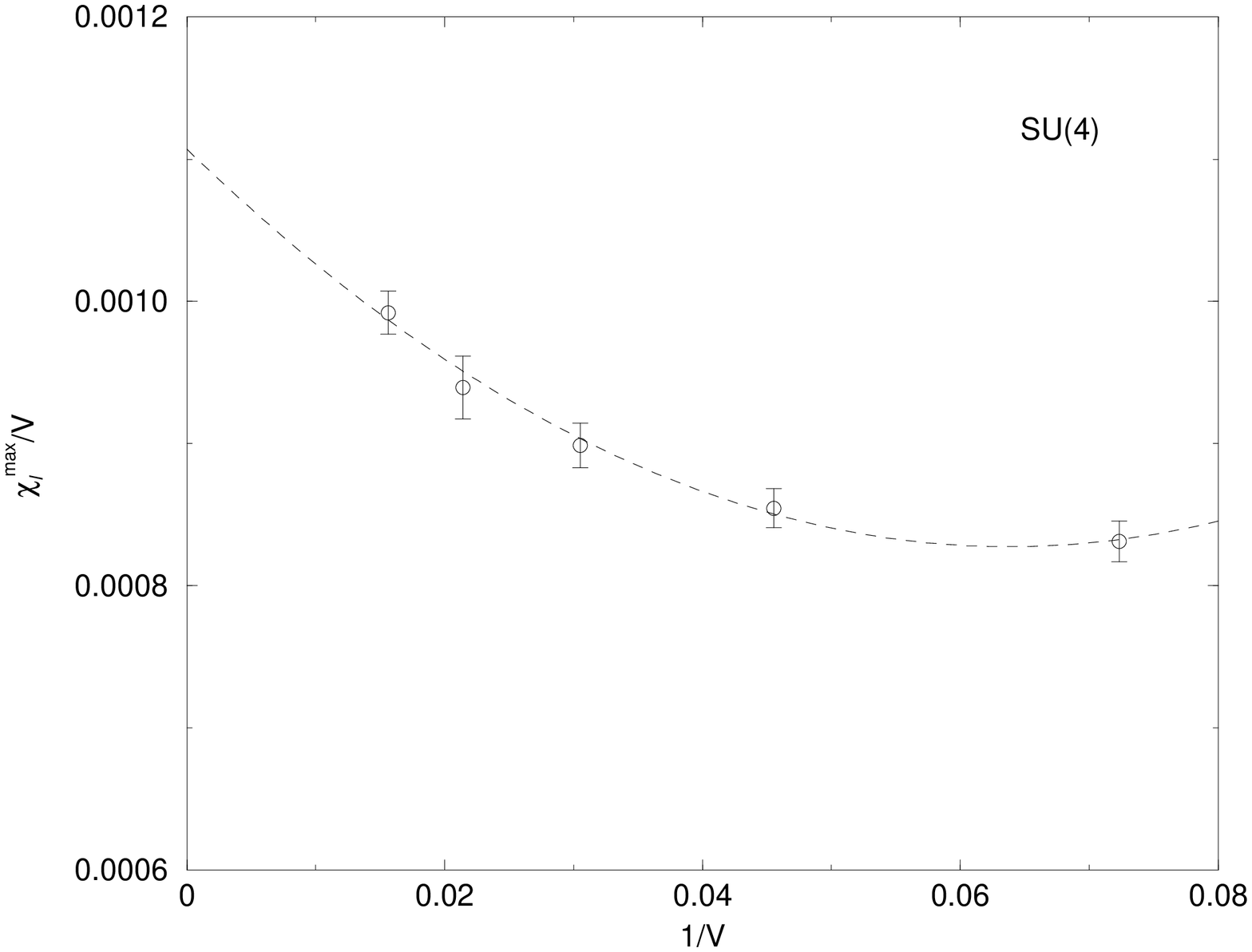}\\
\includegraphics[angle=-90,width=\figsize]{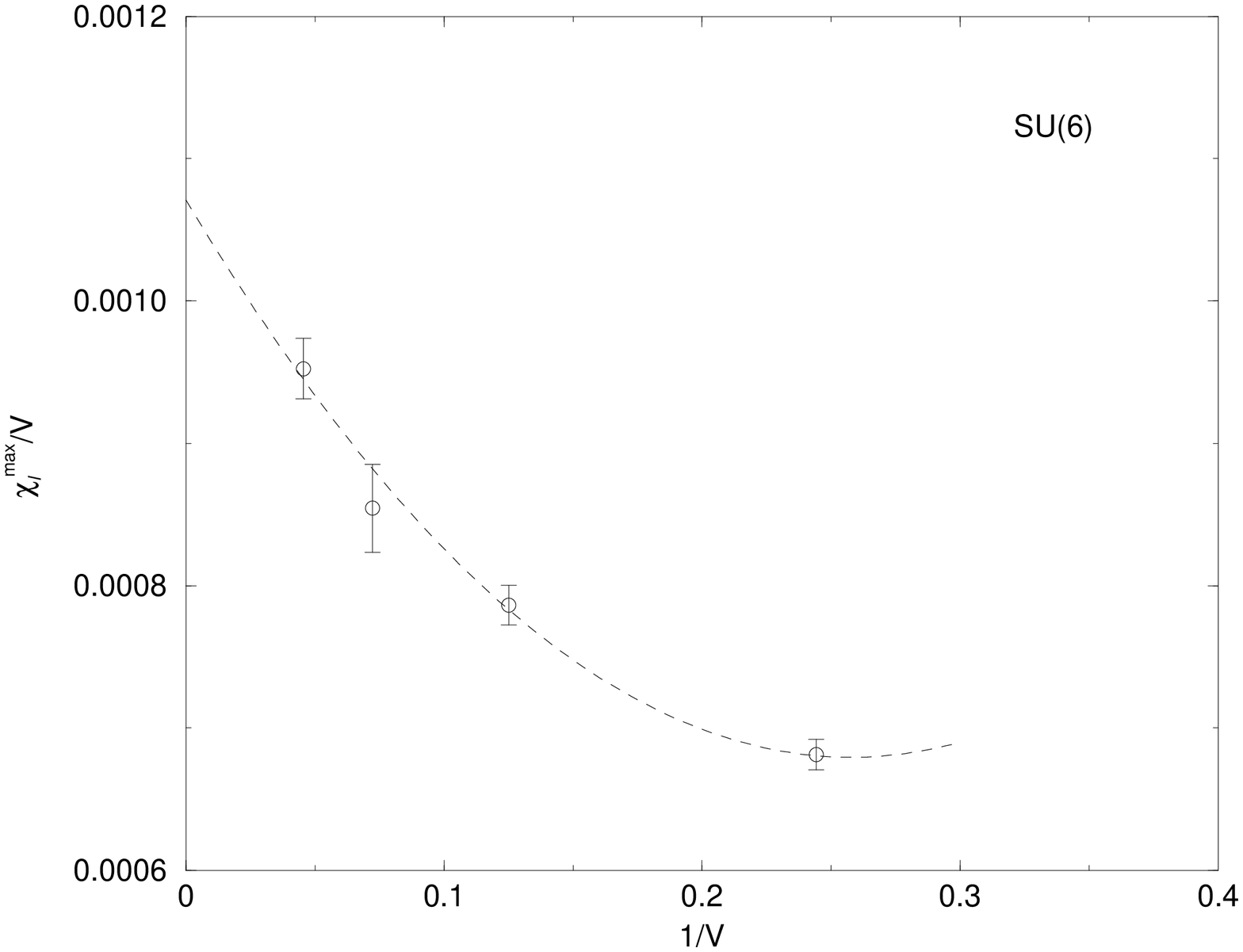}
\end{center}
\caption[]{The normalised susceptibility maxima
$\chi_l^\text{max}/V$ plotted against $1/V$ for $N=4,6$ at
$L_t=5$. The dashed lines are the best fits with $O(1/V)$ and
$O(1/V^2)$ corrections. Note the different scale for SU(6) on the $x$-axis.}
\label{fig:chi_max_scaling_su4_su6_t5}
\end{figure}

\begin{figure}[htb]
\begin{center}
\includegraphics[angle=-90,width=8cm]{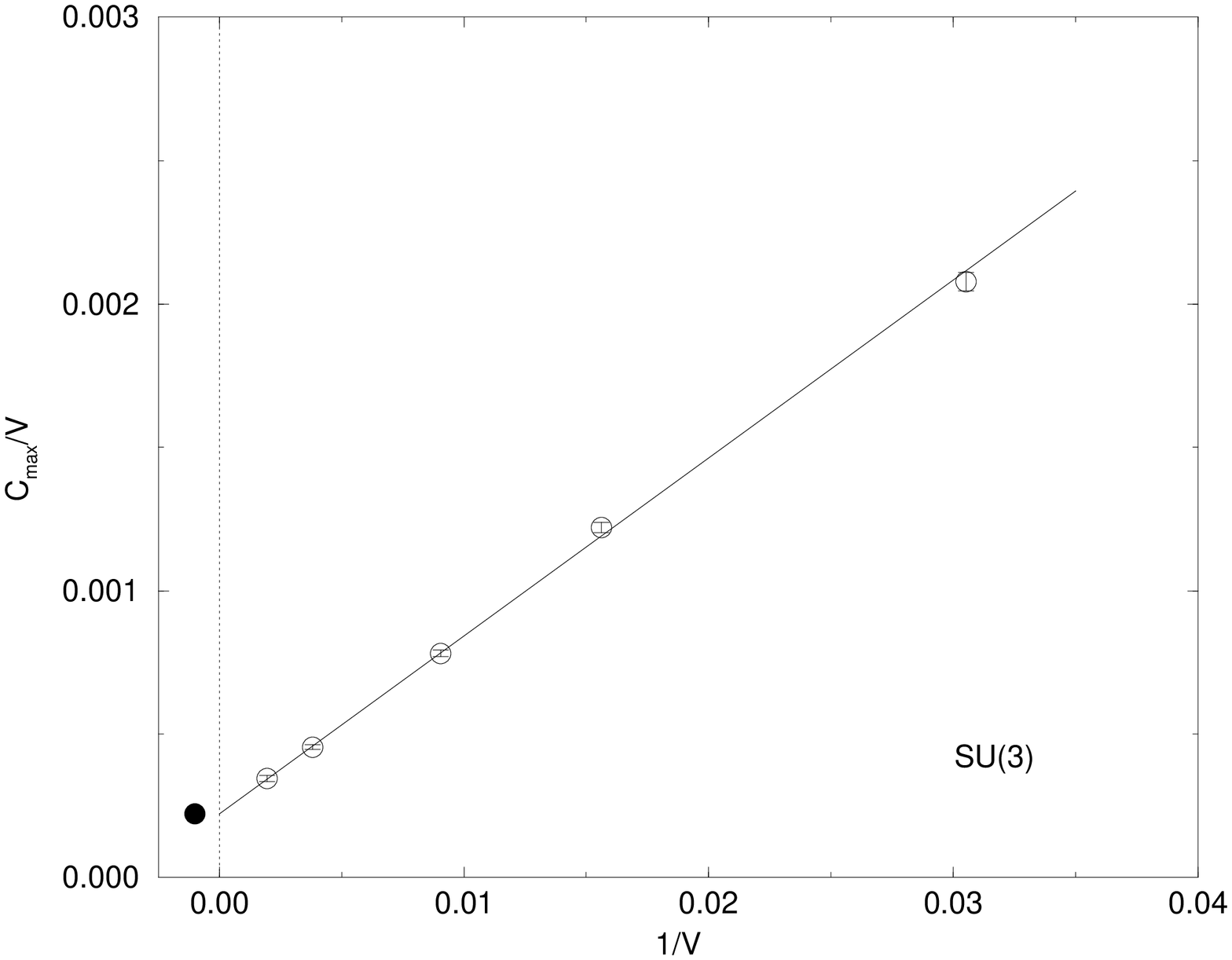}\\
\includegraphics[angle=-90,width=8cm]{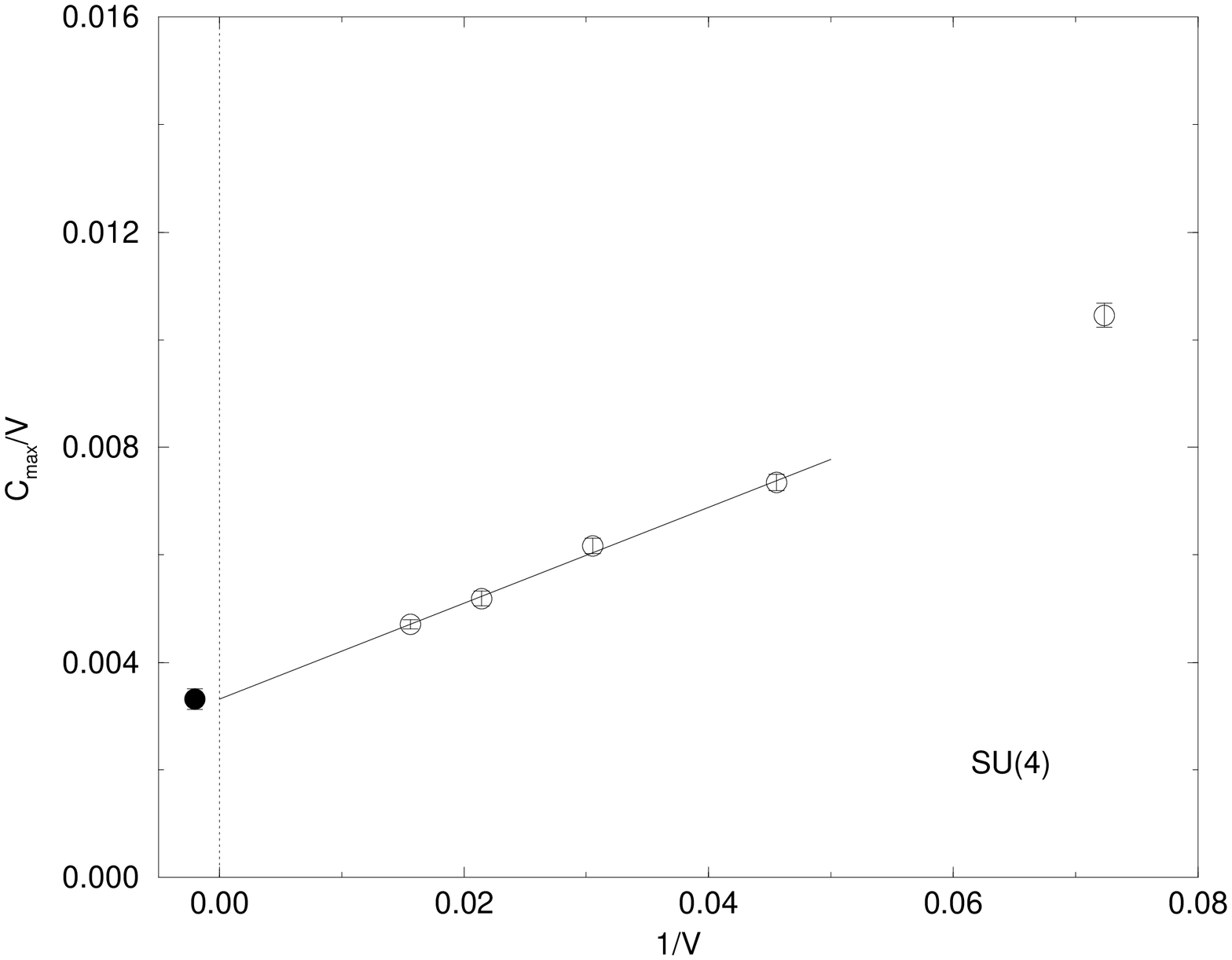}\\
\includegraphics[angle=-90,width=8cm]{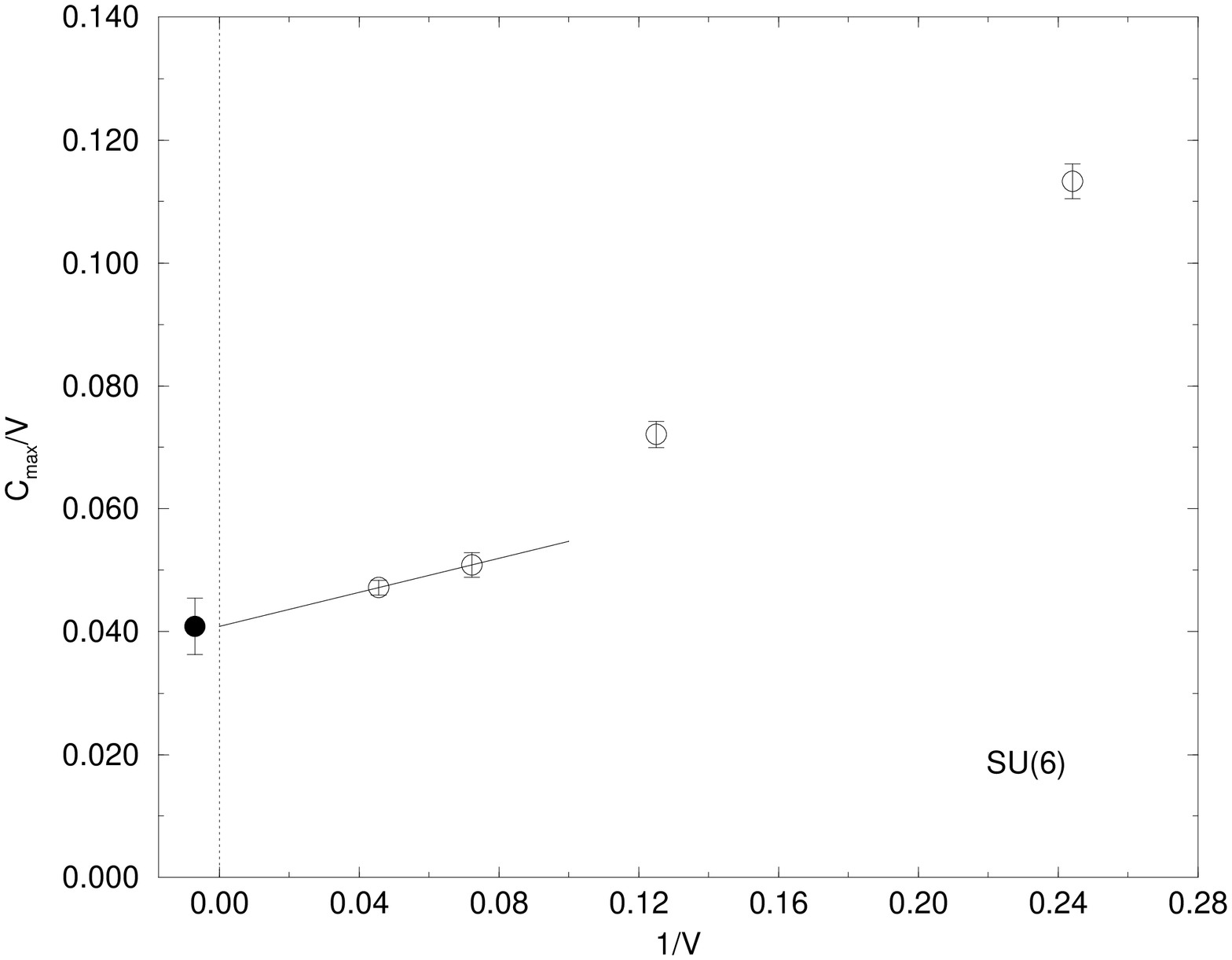}\\
\end{center}
\caption[]{The normalised specific heat maxima $C_\text{max}/V$
plotted against $1/V$ for $N=3,4,6$ at
$L_t=5$. The straight lines are best fits with a leading $O(1/V)$
correction. Also shown are the $V\to\infty$ limits.}
\label{fig:c_max_scaling_su3_su4_su6_t5}
\end{figure}

\begin{figure}[htb]
\begin{center}
\includegraphics[angle=-90,width=\figsize]{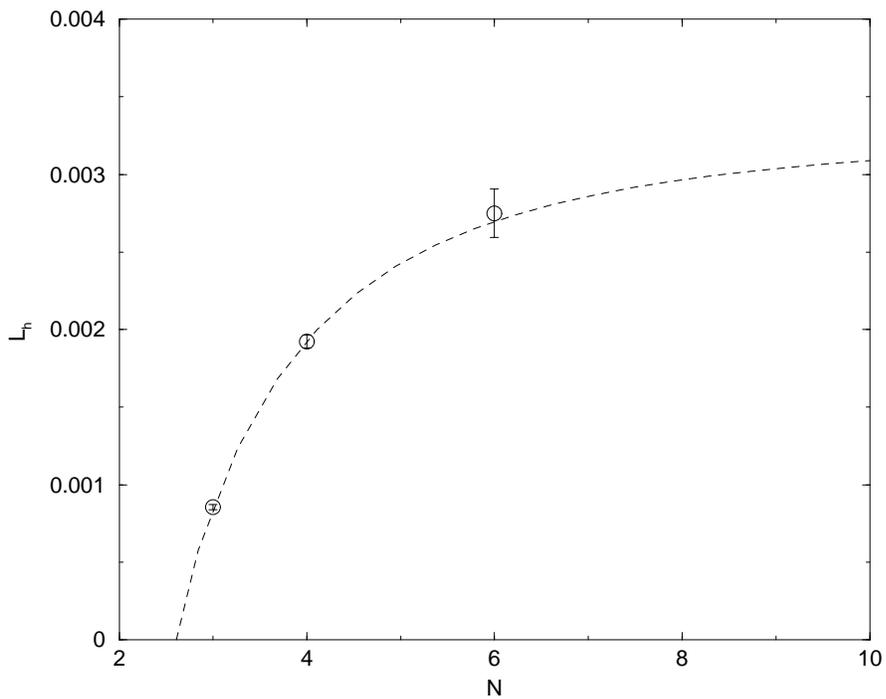} 
\end{center}
\caption[]{The latent heat, $L_h$, plotted versus $N$ at $L_t=5$. The
dashed line is a large-$N$ extrapolation with a leading $O(1/N^2)$
correction.}
\label{fig:latent_heat_vs_N}
\end{figure}

\begin{figure}[htb]
\begin{center}
\includegraphics[angle=-90,width=\figsize]{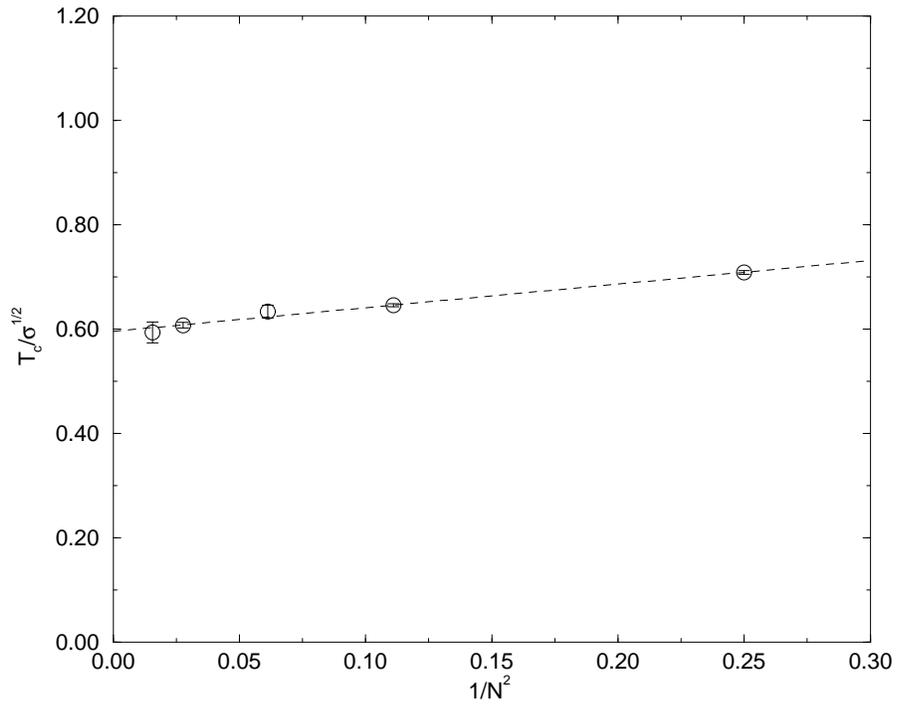} 
\end{center}
\caption[]{The deconfining temperature, $T_c$, in units of the
string tension, $\sigma$ plotted versus $1/N^2$. The
dashed line is the large-$N$ extrapolation with a
leading $O(1/N^2)$ correction.}
\label{fig:tc_sigma_vs_N}
\end{figure}

\begin{figure}[htb]
\begin{center}
\includegraphics[angle=-90,width=\figsize]{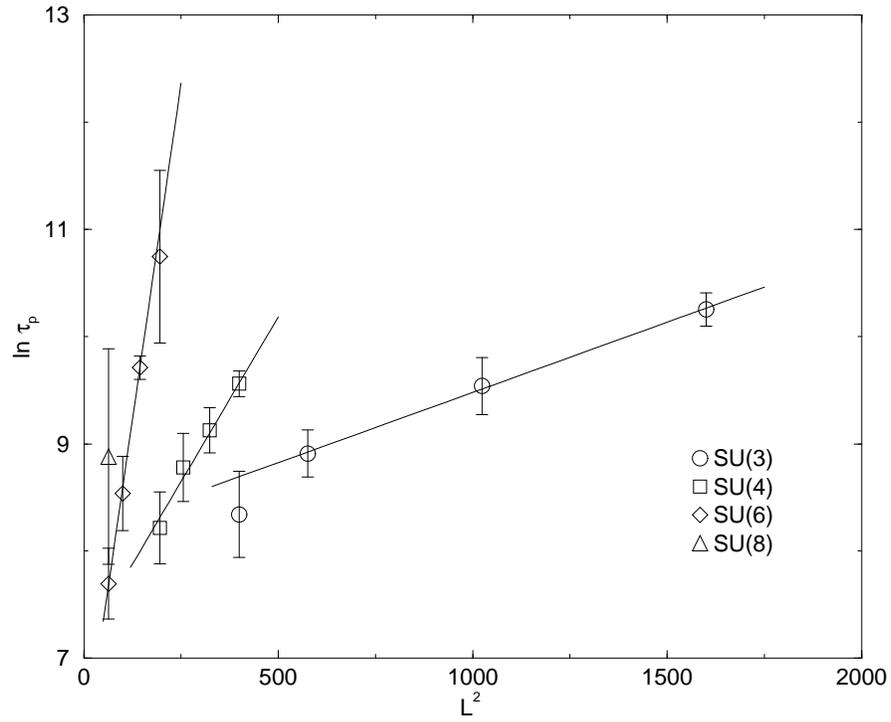} 
\end{center}
\caption[]{The persistence time, $\tau_p$, plotted versus $L^2$ for various
$N$ at $L_t=5$. The straight lines are fits of the form $\tau_p \propto \exp{-c(N) L^2}$.}
\label{fig:persistence_time_sun}
\end{figure}

\end{document}